%% file: main.tex
\documentclass[conference]{IEEEtran}
\IEEEoverridecommandlockouts
\usepackage{cite}
\usepackage{amsmath,amssymb,amsfonts}
\usepackage{algorithmic}
\usepackage{graphicx}

\graphicspath{{figures/}}

\usepackage{textcomp}
\usepackage{xcolor}
\def\BibTeX{{\rm B\kern-.05em{\sc i\kern-.025em b}\kern-.08em
    T\kern-.1667em\lower.7ex\hbox{E}\kern-.125emX}}
  
\begin{document}

\title{Biological Random Walks: integrating heterogeneous data in disease gene prioritization}

\author{
\IEEEauthorblockN{Michele Gentili, Leonardo Martini, Manuela Petti, Lorenzo Farina and Luca Becchetti}
\IEEEauthorblockA{\textit{Department of Computer, Control, and Management Engineering "A. Ruberti"} \\
\textit{Sapienza University of Rome}, Italy \\
\{surname\}@diag.uniroma1.it}

}

\maketitle

\begin{abstract}
This work proposes a unified framework to leverage biological information in network propagation-based gene prioritization algorithms. Preliminary results on breast cancer data show significant improvements over state-of-the-art baselines, such as the prioritization of genes that are not identified as potential candidates by interactome-based algorithms, but that appear to be involved in/or potentially related to breast cancer, according to a functional analysis based on recent literature. 
\end{abstract}


\input{trunk/1_intro.tex}

\input{trunk/2_results.tex}
\input{trunk/5_methods.tex}

\section*{Acknowledgment}

This work was partially supported by  "\textit{Progetti di Ricerca Medi 2018:  Network medicine based machine learning and graph theory algorithms for precision oncology}, id n. RM1181642AFA34C2", and by ERC Advanced Grant 788893 AMDROMA "Algorithmic and Mechanism Design Research in Online Markets" and MIUR PRIN project ALGADIMAR "Algorithms, Games, and Digital Markets"

\bibliographystyle{plain}
\bibliography{main}

\end{document}

%% file: trunk/1_intro.tex
\section{Introduction and related work}
\label{sec:introduction}

Today, big data, genomics, and quantitative {\it in silico} methodologies integration, have the potential to push forward the frontiers of medicine in an unprecedented way \cite{chan2012emerging,gustafsson2014modules}. Clinicians, diagnosticians and therapists have long striven to determine single molecular traits that lead to diseases. What they had in mind was the idea that a single “golden bullet” drug might provide a cure. Unfortunately, individual diseases rarely share the same mutations. This reductionist approach largely ignores the essential complexity of human biology. Indeed, a large body of evidence that is now emerging from new genomic technologies, points out directly to the cause of disease as “perturbations” within the “interactome”,  {\it i.e.} the comprehensive network map of molecular components and their interactions \cite{chan2012emerging}.

As a matter of fact, a growing body of knowledge reveals the association between groups of interacting proteins and disease within the so-called “human interactome”, representing the cellular network of all physical molecular interactions \cite{barabasi2011network}. Precisely, the “human interactome” is composed of direct physical, regulatory (transcription factors binding), binary, metabolic enzyme-coupled, protein complexes and kinase/substrate interactions. Such network is largely incomplete as well as the connections between genes and disease. Currently, more than 140,000 interactions between more than 13,000 proteins are known (see {\it e.g.} \cite{korcsmaros2017next,gustafsson2014modules}. The interactome-based network medicine approach \cite{barabasi2011network} has proved to be very effective in the study of many diseases, {\it e.g.} by identifying putative biomarkers and subtypes to provide a rational approach to drug targeting \cite{barabasi2011network,ozturk2018emerging}.

``Disease proteins'' are the product of genes whose mutations have a causal effect on the respective phenotype. In other words, such proteins work together in a network that gives rise to a cellular function and its disruption ends up in a specific disease phenotype. Disease proteins may provide targets for cancer therapy such as, for example, {\it imatinib} which targets the BCR-ABL fusion or {\it gefitinib} which binds and inhibits EGFR \cite{ozturk2018emerging}. However, the big picture is far more complicated, since a large variety of factors affect the effectiveness of a given drug for a specific patient. For example, targeted inhibition of BRAF V600E in patients harboring this mutation, is very effective in melanoma, but not in colorectal cancer \cite{ozturk2018emerging}. Improvement of precision therapy needs new approaches able to capture information about molecular mechanisms by characterizing disease proteins causing the disruption of tumor driving pathways. 

A key property of the underlying molecular network of interactions is that disease proteins are not found to be uniformly scattered across the interactome, but they tend to interact with one another confined in one or several subgraphs called “disease modules” \cite{menche2015uncovering}. In fact, disease proteins are prone to participate in common biological activities such as, for example, genome maintenance, cell differentiation or growth signaling, which are the most relevant pathways in carcinogenesis \cite{ozturk2018emerging}. Consequently, the “module” property also reflects the biological feature that disease proteins are often localized on specific biological compartments (pathway, cellular space, or tissue). 

These considerations directly point towards the possibility that, whenever a disease module sub-network is found, other disease-related parts are likely to be identified in their topological neighborhood \cite{barabasi2011network}. However, notwithstanding a strong community commitment to find new protein interactions and relevant mutations for disease characterization, the list is largely incomplete. Moreover, identification of specific disease genes is often impaired by gene pleiotropy, by the multi-genic feature of many diseases, by the influence of a plethora of environmental agents, and by genome variability \cite{bromberg2013disease}.

The need for “new” disease genes (or disease proteins) as putative candidates for diagnosis, treatment or drug targeting, motivated the development of a number of algorithms for predicting disease genes and modules \cite{ghiassian2015disease}. The key question is whether it is possible to find a way to fully characterize such genes (with respect to non-disease genes) and find an algorithm able to capture such features. From a network perspective, the goal is to find correlations between disease gene “location” on the interactome and the network topology. In other words, one hypothesizes that disease genes are embedded within modules in ways that are amenable to some topological feature “descriptor”. The recent \cite{menche2015uncovering} evidence-based biological observation that disease genes are not randomly positioned in the interactome has opened new possibilities for developing algorithms for disease gene predictions. 

Two groups of methodologies have emerged in the last decade as the most promising ones: network propagation \cite{cowen2017network} and modules-based \cite{ghiassian2015disease,barabasi2011network} algorithms. Network propagation (or diffusion-based) algorithms rely on the assumption that the “information” contained in the initial (known) set of disease genes, flows through the network through nearby proteins. By contrast, module-based algorithms rely on the hypothesis that all cellular components that belong to the same topological, functional or disease module have a high likelihood of being involved in the same disease. 

From the above discussion, it is clear that prioritizing candidate disease genes using the “interactome”, {\it i.e.} the network of physical protein interactions, and mutational data (known disease gene or “seeds”), is still a largely open problem. A reliable prioritization (or ranking) of new predicted disease genes is very important from a biological viewpoint, since it provides valuable information of a putative specific activity of a gene in the development of a disease. Simply put, the smaller its rank position, the more likely a gene is to be a ''true'' disease one. This allows providing experimenters/clinicians with an ordered list of potentially interesting genes for further scrutiny, possibly speeding the complex and costly task of identifying the most “promising” candidates. 

\noindent\textbf{Our contribution.}
In this work, we provide a unified framework to leverage biological information in network propagation-based gene prioritization algorithms. This brings to significant improvements over state-of-the-art algorithms. In more detail, we modify a well-known random walk-based, flow propagation algorithm \cite{kohler2008walking}, modifying the dynamics of flow propagation according to the functional relevance of nodes for the disease under consideration. We considered the same diseases as \cite{ghiassian2015disease} and multiple biological data sources. In the remainder however, for the sake of space and for clarity of exposition, we focus on  \textit{breast cancer} as a use case and we adopt \textit{Gene Ontology Annotations - biological process} \cite{DBLP:journals/nar/Consortium19a} (GO in the remainder) as added biological information. 

In the output ranking of the algorithm we almost double the number of known disease genes in the first 50 positions with respect to state of the art baselines, in particular DIAMOnD \cite{ghiassian2015disease} and Random Walk with Restart \cite{kohler2008walking}. Moreover, some very promising candidates are prioritized by our algorithm but not by baselines. Of these, some  were only recently associated to breast cancer, while others appear to be potentially related to the disease according to a functional analysis based on recent literature, as discussed in Section \ref{sec:results}.

\noindent\textbf{Roadmap.} The rest of this paper is organized as follows. Section \ref{sec:results} gives an overview of our findings and their potential biological relevance. Section \ref{sec:methods} provides  background about the baselines we considered, a more detailed account of our approach and of the experimental setting. Due to space limitations, it was only possible to include the most significant results and a concise report of experimental evidence supporting the design choices we made.

\begin{figure*}

\centering
\includegraphics[width=0.7\paperwidth]{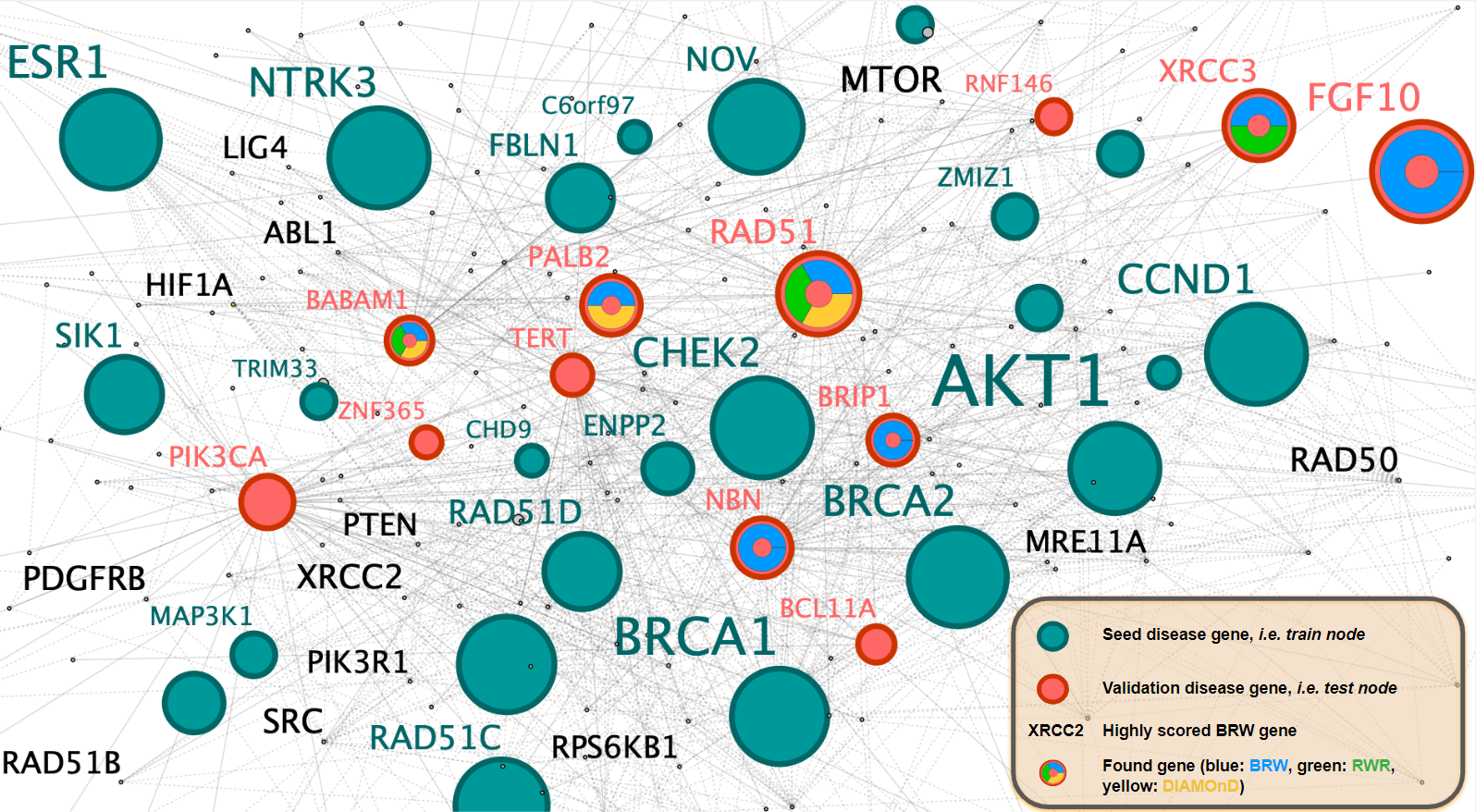}

\caption{\textbf{PPI single experiment result.} The Figure shows: the splitting of the known disease genes, \textit{i.e.} train (seed) and test (validation) nodes, the retrieved nodes by the different algorithms and the nodes' names that have been highly ranked by the BRW. The size of train and test nodes is proportional to the BNR, indeed node \textbf{\textit{FGF10}} has a very high score, inducing a high ranking despite has only one connection. Interestingly, BRW finds nodes retrieved by RWR and DIAMOnD algorithms. Furthermore, the names of other highly ranked nodes by the BRW that aren't present in the test set are shown, and they seem to be promising candidate as breast cancer related genes, such as \textbf{\textit{RAD50}} \cite{heikkinen2006rad50}, \textbf{\textit{XRCC2}} \cite{park2012rare}.}

\label{fig:network}
\end{figure*}


%% file: trunk/2_results.tex
\section{Results and discussion}\label{sec:results}

In this section, we present the main findings of our work. In particular, we discuss the benefits of leveraging both biological and interactome information within gene prioritization algorithms. To this purpose, we compared state of the art prioritization algorithms that only rely on analysis of the interactome, namely DIAMOnD \cite{ghiassian2015disease} and Random Walk with Restart \cite{kohler2008walking}, with two heuristics we propose: i) Biological Node Relevance (BNR) only leverages biological information (e.g., annotations) to prioritize genes; ii) Biological Random Walk (BRW) is a random walk-based heuristic that, differently from \cite{kohler2008walking}, also leverages biological information to bias the random walk toward genes that are functionally closer to known disease genes according to current literature. 

BRW builds on the hypothesis that integrating different biological information sources may better reflect the complexity of protein interactions in a cell's process. In light of this insight, our algorithm integrates information on pairwise protein interaction reflected in the Protein-Protein Interaction network (PPI) \cite{barabasi2011network} with other biological data in a unified framework. Our approach is to some extent agnostic to the particular biological data source, as long as it affords a principled notion of similarity between proteins. So for example, while we focus on gene annotation data in this paper, the same approach can be adopted to integrate different sources of biological information, {\it e.g.} miRNA targets or pathway annotation data.


\subsection{The role of biological information}
We began by investigating the potential role of high-quality, biological information (gene annotations in this case) in prioritizing new candidates genes. To this purpose, we designed a very simple heuristic that ranks genes of the PPI network according to the degree of their co-occurrence in biological processes, completely disregarding mutual interaction properties encoded by the PPI itself. Our \textbf{Biological Node Relevance} heuristic (BNR in the remainder) prioritizes genes only on the basis of their functional similarity with a seed set of known disease genes, with similarity measured on the basis of annotation data from the Gene Ontology database \cite{DBLP:journals/nar/Consortium19a} according to well-established similarity indices  adopted in Data Science. While details are provided in Section \ref{sec:methods}, a high-level description of the BNR is given in the paragraphs that follow.

\noindent Given a seed set $S_0$ of known disease genes, BNR ranks new candidate genes according to the following steps:

\begin{enumerate}
\item We first compute the set of statistically significant annotations for genes belonging to the seed set $S_0$. We call this the \textit{enriched set} of annotations for the disease.
\item BNR then ranks each node \text{i} according to its \text{biological relevance} BNR(i), namely, the extent of the overlap between the \textit{enriched set} and the set of gene's annotations.
\end{enumerate}

Whilst it is reasonable to expect that curated, high-quality annotations are likely to contain information that can be leveraged to the purpose of gene prioritization, we observed that BNR outperforms state of the art topology and flow-based prioritization heuristics, in particular DIAMOnD \cite{ghiassian2015disease} and the well-established diffusion method based on random walks with restart \cite{kohler2008walking}. In particular, as shown in Figure \ref{fig:brw}, BNR consistently recovers a larger fraction of known disease genes among its top-$k$ ranking candidates. This result suggests that biological annotations (and, hopefully, other curated data) contain rich information, which is not implicit in PPI networks and thus cannot be leveraged by standard topology or flow-based methods.

\subsection{A unified framework}
The \textbf{Biological Random Walk} (BRW in the remainder) heuristic provides a framework to integrate heterogeneous biological data sources within diffusion-based prioritization methods that are based on the well known Random Walk with restart algorithm (RWR).
For the sake of exposition, in the remainder we refer to the biological information associated to a gene \textit{i} (e.g., the set of its annotations) as the set of its \textit{labels}, denoted by \textit{labels(i)}. In this study, we only used annotations from the Gene Ontology (GO in the remainder) database to define labels, since at the moment it is one of the most complete and best curated available datasets. We remark however, that in principle any reliable information source on gene biology can be integrated.
BRW ranks genes according to the following steps:

\begin{enumerate}
\item We compute the set of statistically significant annotations of known disease genes, as for the BNR heuristic, {\it i.e.}, the \textit{enriched set}
\item Rather than using the standard method\footnote{Whilst details are given in Section \ref{sec:methods}, here we remind that in the standard RWR approach \cite{kohler2008walking}, the probability of restarting the random walk from a given seed node (disease gene) is the same for all seeds nodes, while it is $0$ for other nodes of the PPI.}, we compute individual teleporting probabilities for all nodes of the PPI. In particular, the \textit{Biological Teleporting Probability (BTP)} of a node increases with the similarity between its labels and the \textit{enriched set} (details in Section \ref{sec:methods}), 
\item In a similar fashion, we weigh PPI network interactions using node annotations and the \textit{enriched set}. This results in a modified random walk, namely the \textit{Biological Random Walk (BRW)}, in which flow propagation is biased toward genes that are functionally closer to those forming the seed set.
\item Finally, we rank genes according to their \textit{Biological Random Walk (BRW)} score.
\end{enumerate}

As the example in Figure \ref{fig:brw} highlights, BRW not only propagates flow to and from known disease genes, but also involves a broader set of genes that are functionally related to disease ones, though themselves not directly related to the disease, at least to the best of our knowledge.

\begin{figure}
\centering
\includegraphics[width=0.9\linewidth]{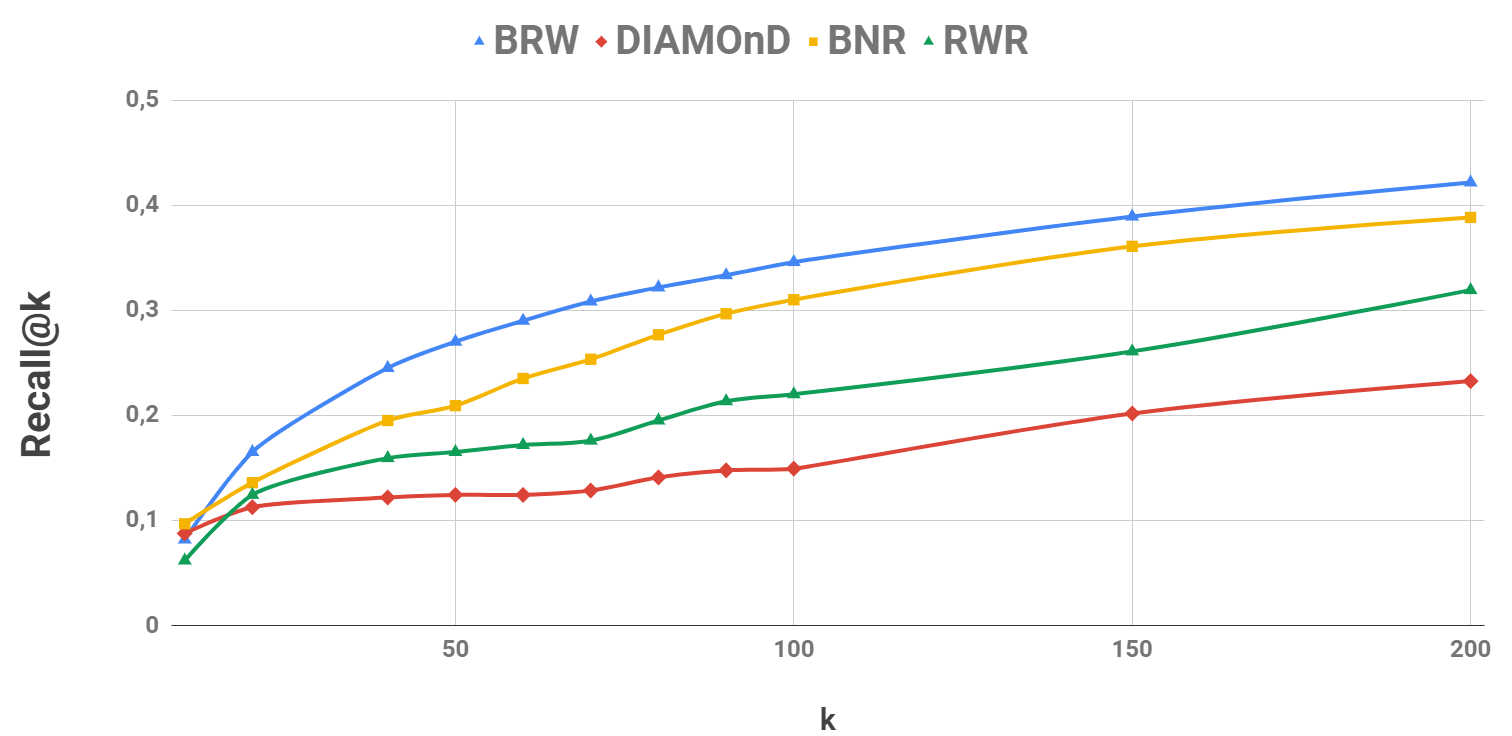}
\caption{\textbf{Recall @k scores} show how BRW performs better than other state-of-the-art techniques. The recall is computed splitting the known disease genes into two groups the seed nodes and the test nodes. The former are used to run the algorithm, the latter are used has validation for the output of the algorithms. The recall@k are the percentage on nodes of the test set discovered in the first k position of the ranked lists, that are the output of the algorithm. The splitting of the genes has been repeated 100 times, and the score is the average of the experiment results. The BRW on average is able to find more than $40\%$ of the test nodes in the first 200 ranked genes.}
\label{fig:recall}
\end{figure}

The results of Figure \ref{fig:recall} suggest that BRW seems to leverage both heterogeneous sources of biological information. In particular, it significantly outperforms RWR and DIAMOnD, but it also achieves better recall than the BNR baseline across the entire spectrum of the values $k$ that we considered. We also note that best results are achieved using a value $0.75$ for the restart probability. This intuitively means that best candidates are mostly found in the vicinity of disease genes or genes that are functionally related to them.

Beyond this internal validation of a more quantitative nature, the paragraphs that follow report and discuss anecdotal evidence, as to the potential biological interest of candidate genes that are prioritized by our algorithm, but are not part of the pool of known disease ones. 

\subsection{Functional analysis}

\begin{table*}[]
\caption{Prioritized genes found only by BRW algorithm and not by other, using as initial disease genes (\textit{seed nodes}) all the known genes. Details on PPI and algorithms in Section \ref{sec:methods}.The left columns indicates the parameters used to define the enriched set.}
\label{tab:res}
\centering
\begin{tabular}{|c|c|c|l|}
\hline
\textit{Benjamini-Hochberg Correction} & \textit{P-value} & \textit{Annotations in Enriched set} & \textit{BRW prioritized genes}\\
\hline
 False & 0.01 &213&  ERBB4, ERBB2, CDC42, TGFB1, FGF9, SIRT1 \\
 \hline
False & 0.05 & 485 &  ERBB4, VEGFA, BMP4, TGFB1, BMP2,	FGF9  \\
\hline
True & 0.01 &  80& ERBB4, TGFB1, BMP4, CDC42\\
\hline
 True & 0.05 & 318&  ERBB2, PAK1, ERBB4, RAD50, XRCC2 \\
 \hline
\end{tabular}
\end{table*}

Table \ref{tab:res} reports genes prioritized by our BRW algorithm only. Therefore, we briefly discuss the relevance of some of them to breast cancer, which is the most common malignancy in women \cite{assi2013epidemiology} and has the second highest incidence among all types of cancer worldwide. Notably, the list in table X contains two members of the erbB family which is composed of closely related genes: erbB (her), erbB-2 (her-2, neu), erbB-3 (her-3), and erbB-4 (her-4). This genes also encode members of the epidermal growth factor (EGF) receptor family of receptor tyrosine kinases. In particular, erbB-2 gene is a proto-oncogene. In fact, overexpression of ErbB-2 leads to transformation, tumorigenicity, and metastasis. These findings support the implications of ErbB-2 as a major player in breast cancer initiation and/or progression. Moreover, targeting of ErbB-2 has proved to be effective for drug development \cite{stern2000tyrosine}. Over expression of human epidermal growth factor receptor-2 (ErbB-2) has been found in 20-30\% of breast cancer patients and widely recognized as a reliable marker for metastatis formation, drug resistance and high aggressiveness. Among all of the drugs that target HerbB-2, trastuzumab, pertuzumab, trastuzumab emtansine and lapatinib have been proven to be effective in several clinical trials \cite{verma2012trastuzumab}. Another important gene in our list is vegfa, a member of the Vascular Endothelial Growth Factor (VEGF) family which plays an important role in multiple physiologic and pathologic processes involving endothelial cells. Several preclinical and clinical evidence supports its relevance in breast cancer and, consequently, numerous anti-VEGF drugs are now being under clinical evaluation \cite{sledge2005vegf}. Interestingly, gene fgf9 of our list, plays a role in many tumours, like breast cancer, that contain different populations of cells which may show increased resistance to anticancer drugs. There are now evidences of "cancer stem-like cells, which are important for survival and expansion of normal stem cells. It has been reported that, in analogy to embryonic mammary epithelial biology, estrogen signaling expands the pool of functional breast cancer stem-like cells through a paracrine Fgf/Fgfr/Tbx3 signaling pathway \cite{fillmore2010estrogen}. Moreover, bmp4 gene in our list, encode the bone morphogenetic protein 4, which is a key regulator of cell proliferation and differentiation. In breast cancer cells, bmp4 is able to reduce proliferation and induce migration, invasion and metastatis formation in vitro \cite{ampuja2016impact}. Last (but not least) we found gene p63 in our list which is a transcription factor of the p53 gene family, widely known to play a fundamental role in the development of all the stratified squamous epithelia, including breast \cite{di2016p63}.

\subsection{Robustness}
We briefly mention here the robustness of our results to the presence of possible noise in both interactome and annotation data, finding that our framework is resilient to degree preserving random shuffling on the graph\cite{milo2003uniform} and it partially decreases its performances when noising the annotation.

\begin{figure}[bhp]
\includegraphics[width=\columnwidth]{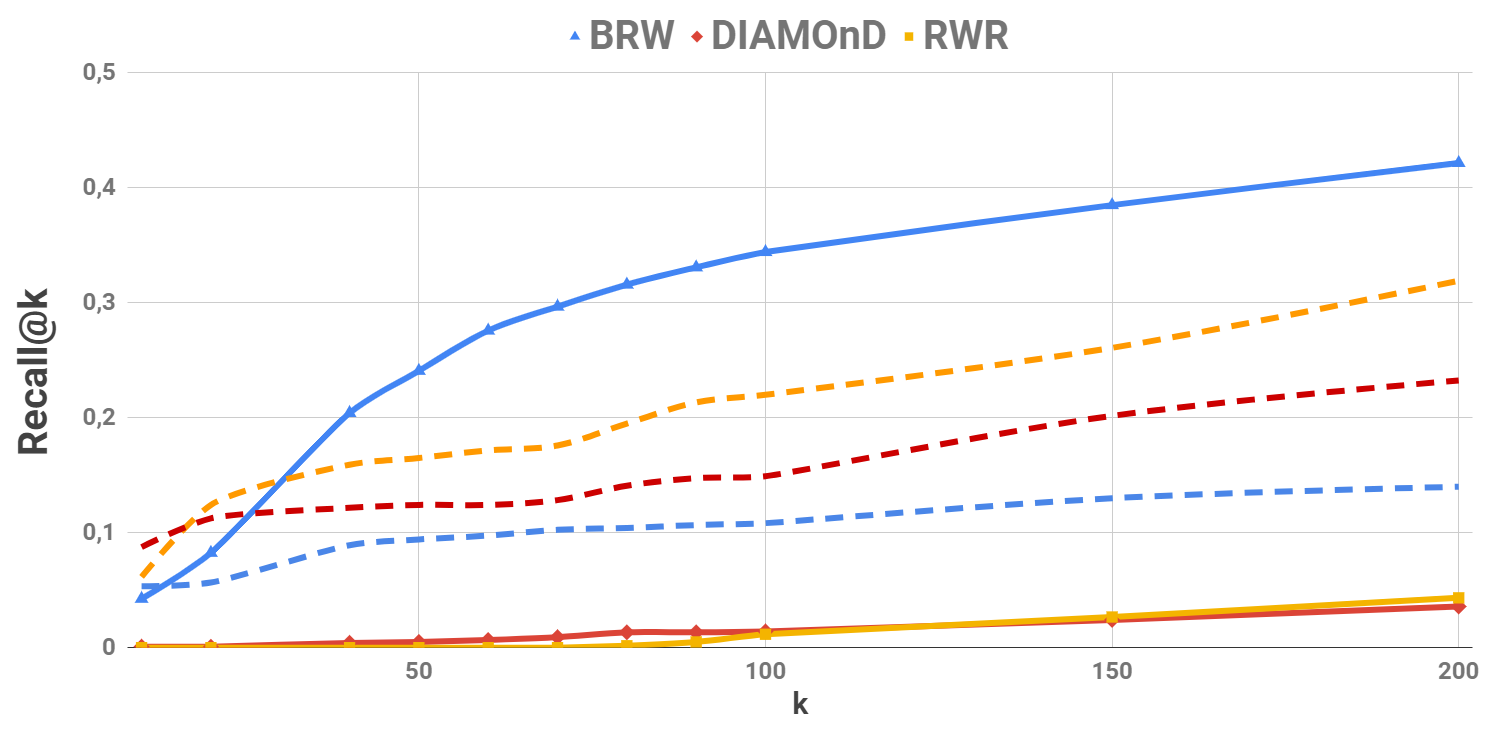}

\label{fig:networknoise}
\caption{\textbf{PPI robustness analysis:} Recall@k is measured after shuffling the PPI interactions keeping node's degree (\textit{continuous lines}), and after shuffling node's annotations (\textit{dashed lines}). As we can see all algorithms that use only topological information fail in the case their input is noisy. BRW decrease by 2/3 when shuffling the annotations: relying on two different sources, the BRW is more resilient to noise.}
\end{figure}

%% file: trunk/5_methods.tex
\section{Materials and methods}
\label{sec:methods}

\subsection{Datasets}
\subsubsection{PPI Network and Gene-Disease Associations}
The experiments discussed in Section \ref{sec:results} were conducted on the same PPI network as \cite{ghiassian2015disease} for the sake of comparison. In \cite{ghiassian2015disease}, the authors only considered direct physical protein interactions with reported experimental evidence. Several data sources were used to derive this PPI network:
\begin{itemize}
    \item 	TRANSFAC\cite{kel2003transfac}: this database lists regulatory interactions derived from the presence of a transcription factor binding site in the promoter region of a certain gene;
    \item IntAct\cite{Armean2009IntAct}, MINT\cite{Aryamontri2009mint}, BioGRID\cite{aryamontri2010Biogrid} and HPRD\cite{Venugopal2008HPRD}: these databases list physical PPI interactions, typically identified by low throughput experiments and  manually curated by experts;
    \item KEGG and BIGG\cite{Schellenberger2010BIGG}: sources used to find metabolic enzyme-coupled interactions;
    \item CORUM\cite{Ruepp2009CORUM}: this database lists mammalian protein complexes as single molecular units that integrate multiple gene products.
\end{itemize}

\noindent In addition, we considered the main connected component of the network and we removed self-loops (i.e., edges describing proteins' self-interactions). The resulting graph consists of 13396 nodes and 138405 edges.

\begin{figure}
\centering
\includegraphics[width=0.9\linewidth]{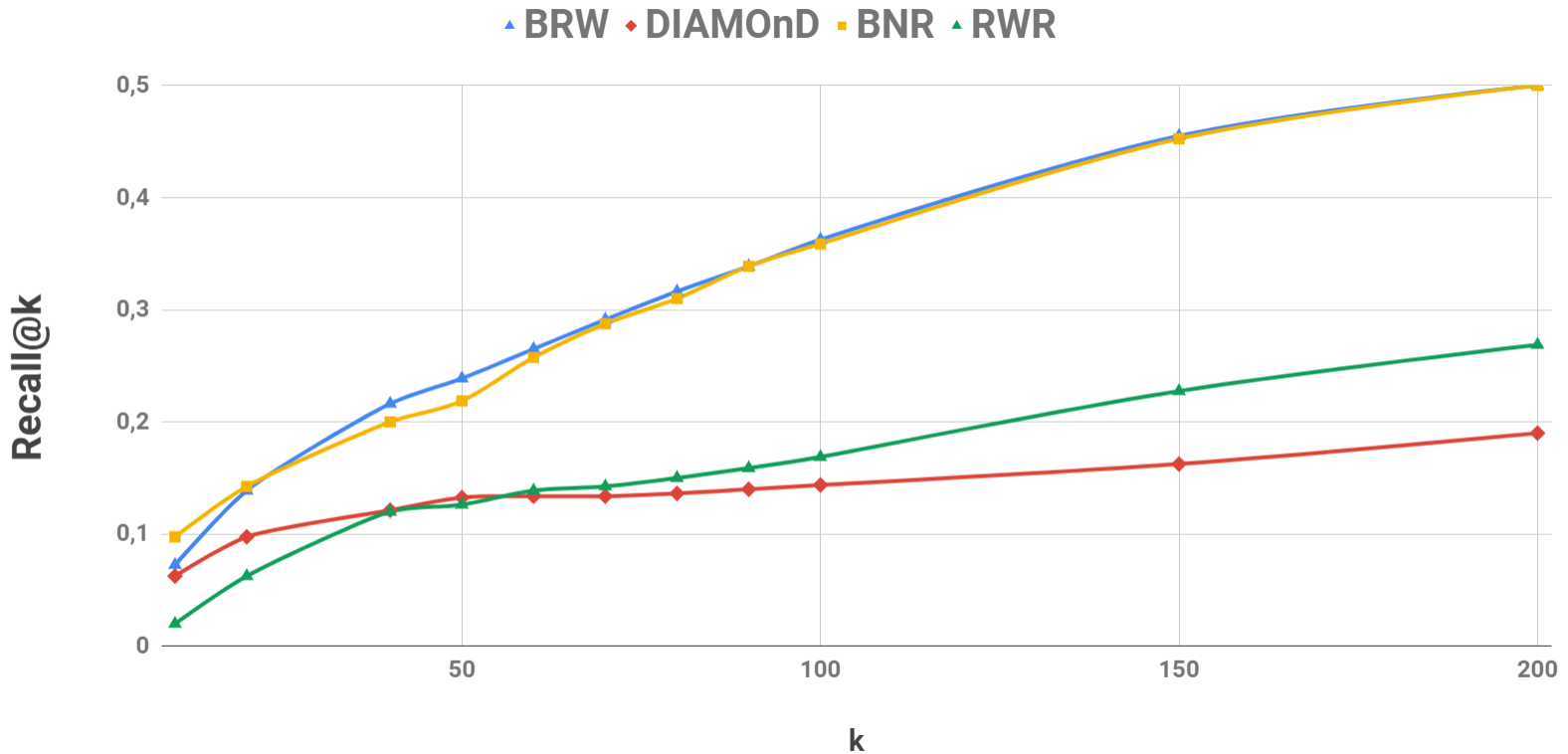}
\caption{\textbf{Recall @k scores} on HIPPIE network \cite{alanis2016hippie}.}

\label{fig:recall_hippie}
\end{figure}

\noindent Disease genes association are the same as in \cite{ghiassian2015disease}. Out of a corpus of 70 diseases in which gene-disease associations were retrieved from OMIM (Online Mendelian Inheritance in Man \cite{OMIM}), in this work we focus on the \textbf{Breast Cancer} phenotype, which involves 40 genes, refer to \cite{ghiassian2015disease} for the complete list. Experiments concerning other diseases will be described and discussed in the journal version of the paper. Though, repeating the same experiment on a different PPI, \textit{HIPPIE} \cite{alanis2016hippie}, we obtain coherent results, see Figure \ref{fig:recall_hippie}.

\subsubsection{Gene annotations}\label{subse:annotations}
We retrieved gene biological information from \textbf{Gene Ontology Consortium: in this case, we extracted annotations describing genes' biological processes. We downloaded the database in November 2018.}

\medskip

\subsection{Algorithms}
In the remainder, we use bold lowercase to denote vectors and capital, non-bold letters to denote matrices. Given a vector $\mathbf{x}$, $x_i$ denotes its $i$-th entry. We use $S$ to denote the subset of PPI's nodes associated to known disease genes, i.e., what we call the \textit{seed set}.
\subsubsection{Random Walk with Restart}\label{subse:rwr}
Random Walk with Restart (RWR) \cite{kohler2008walking} is a diffusion-based method, whose purpose is identifying pathways that are topologically ``close'' to known disease genes in the interactome. It was shown to outperform other prioritization algorithms in many cases \cite{Navlakha}. 

In a nutshell, this algorithm can be seen as performing multiple random walks over the PPI network, each starting from a \textit{seed node} associated to a known disease gene, iteratively moving from one node to a random neighbour, thus simulating the diffusion of the disease phenotype across the interactome. More formally, the random walk with restart is defined as:
\begin{equation}\label{eq:rwr}
        \mathbf{p}^{(t+1)} = (1 - r) W \mathbf{p}^{(t)} + r \mathbf{q}.
\end{equation}
Here, $W$ is the column-normalized adjacency matrix of the graph and  $\mathbf{p}^{(t)}$ is a vector, whose $i$-th entry $p_i^{(t)}$ is the probability of the random walk being at node $i$ at the end of the $t$-th step. $r\in (0, 1)$ is the restart probability. It is the probability that the random walk is restarted from one of the (disease-associated) seed nodes in the next step. Upon a restart, the probability of restarting the random walk from some seed node $j$ is $q_j$. This random walk corresponds to an ergodic Markov chain \cite{levin2017markov} that admits a stationary distribution (i.e., a fixed point) $\mathbf{p}$. Nodes of the PPI are simply ranked by considering the corresponding entries of $\mathbf{p}$ in descending order of magnitude.

Following \cite{kohler2008walking}, in our implementation, the initial probability vector $\mathbf{q}$ was uniform over the subset of seed nodes, i.e., $q_j = 1/|S|$ if $j\in S$, $q_j = 0$ otherwise. We considered the following values for the restart probability: $r\in\{0.25, 0.50, 0.75\}$.\footnote{Note that \cite{kohler2008walking} only considered the  value $0.75$.}

\subsubsection{DIAMOnD Algorithm}

The DIAMOnD (Disease Module Detection) algorithm \cite{ghiassian2015disease} relies on the hypothesis that disease associated proteins do not necessarily reside within locally dense communities. Instead, this algorithm identifies connectivity significance (see paragraphs that follow for definition) as the most predictive quantity.  DIAMOnD exploits this quantity to identify the full disease module starting from  a seed set of known disease proteins.

Consider a PPI network of $N$ nodes, out of which a subset $S$ of seed proteins are associated with a particular disease. Now, consider a protein with $k$ links in the PPI network, out of which $k_s$ to seed nodes. If seed proteins were distributed uniformly at random in the network (null hypothesis), the probability $p(k,k_s)$ that a protein with a total of $k$ links has exactly $k_s$ links to seed proteins (connectivity significance) would be given by the hypergeometric distribution:
\[ p(k,k_s) = \frac{\binom{|S|}{k_s} \cdot \binom{N - |S|}{k - k_s}}{\binom{N}{k}}
\]
To evaluate whether a certain protein has more connections to seed proteins than expected under this null hypothesis, the DIAMOnD algorithm computes its connectivity p-value.



We followed the implementation of DIAMOnD. Note that the set of p-values has to be recomputed in each iteration, which makes the algorithm computationally demanding for moderately large values of $k$. In our experiments, we considered values of $k$ up to $200$.

\begin{figure*}

\centering
\includegraphics[width=0.9\paperwidth]{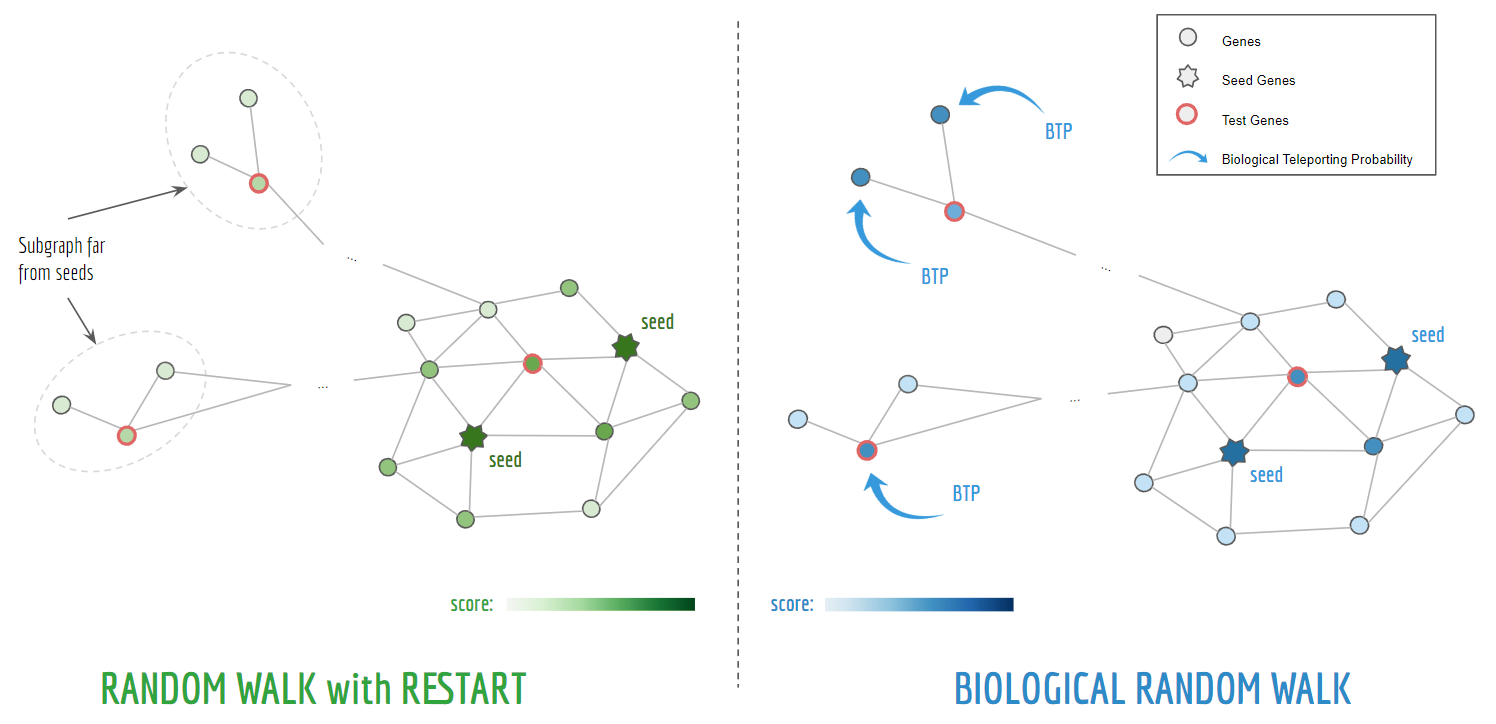}

\caption{\textbf{Biological Random Walk} flow propagation: given the seed nodes (\textit{star nodes}), the flow propagates to his neighbors. The BRW not only propagate the flow around them but also teleports the flow to the target of the BTP nodes (\textit{blue arrows}). So it discovers nodes that are biologically correlated to the seed nodes (just through the BTP, left-lower test node) and those nodes that aren't reached directly to the BTP but are close to many related nodes, the BRW node (\textit{left-upper test node})}
\label{fig:brw}
\end{figure*}

\subsubsection{Biological Node Relevance}
Building on the hypothesis that genes involved in the same disease tend to be functionally related and thus share similar biological information, we came up with a very simple heuristic that ranks genes in the PPI network according to the extent of their co-occurrence within biological processes, what we call Biological Node Relevance (BNR) henceforth. BNR is a simple (yet effective as we shall see) baseline, which completely disregards information implicit in the link structure of the PPI network.

\medskip

\noindent Given a set $S$ of seed nodes (known disease genes), the algorithm first computes the set of annotations (see Section \ref{subse:annotations}) that are statistically significant for seed proteins, \textit{i.e.} the \textit{enriched set},\footnote{We again remind that, while we consider gene annotations here, the same approach can be adapted to different biological data.}, by using Fisher’s exact test to this purpose, with the \textit{P-value} equal to 0.05 and the \textit{Benjamini-Hochberg} correction.

\medskip
\noindent Next, given a list of $N$ proteins  (nodes of the PPI network), BNR computes the score of each protein $i$, defined as the intersection between the set of annotations in the enriched set  and the biological information of $i$, i.e.:

\[
    score(i) = |enriched \_ set \cap label(i)|,
\]
where $label(i)$ is the set of annotations of protein i.

\medskip
\noindent Finally, BNR sorts proteins in descending order with respect to their scores.

\subsubsection{Biological Random Walk}
The Biological Random Walk (BRW) is a framework that exploits both biological (GO annotations, KEGG pathways and miRNA) and topological information (PPI network) to uncover potentially new disease genes.

BRW is essentially a random walk with restart algorithm. While the form of the governing equation is still \eqref{eq:rwr}, the key differences are that both the transition matrix $W$ and the restart vector $\mathbf{q}$ now depend on available genes' biological information. For this reason, we call $W$ and $\mathbf{q}$ respectively \textbf{Biological Transition matrix} and \textbf{Biological Teleporting Probability} vector in the remainder of this section. 

Note that, differently from RWR, $\mathbf{q}$ and $W$ also depend on available biological information, so that the stationary distributions (and thus the rankings) produced $RWR$ and $BRW$ generally differ.

Since the biological relevance of a node can't be used straight forward as a probability, we next describe how we generate $\mathbf{q}$ and $W$, we explored several possibilities for integrating and factoring in available biological information. This implies the setting of several parameters. We used a grid search to select the parameter configuration used in the final round of experiments. For the sake of space, we refer to the case of breast cancer as a disease and GO annotations as complementary (with respect to the PPI) biological information. The approach applies seamlessly to other data sources, such as miRNA or KEGG (results will appear in the journal version of the paper). In the remainder, $labels(i)$ denotes the set of annotations associated to a node $i$ of the PPI network. As usual $S$ denotes the seed set of known disease nodes.

\paragraph{Biological Teleporting Probability (BTP) vector}
The $i$-th entry $q$ of the BTP vector is defined as follows:
\begin{enumerate}
    \item A measure of the overlap between $labels(i)$ and $enriched\_set$ is computed. We call this the \textbf{Node Relevance} $NR(i)$. In this work we considered the following definitions:
    \begin{itemize}
        \item $NR(i) = \frac{|enriched \_ set \cap label(i)|}{|enriched \_ set|}$,
        \item $NR(i) = 1$, whenever $i\in S$.
    \end{itemize}
    \item We let $w_i = \min\{t, f(NR(i))\}$, with $f:\mathbb{R}\rightarrow [0, 1]$ a suitable monotonically increasing function (\textbf{Node Relevance Function}), and $ t \in [0,1]$ used as a parameter to weight the importance of the \textit{NR} score overall. A number of possible choices for $f$ are presented in the paragraphs that follow.
    \item $q_i = w_i/(\sum_i w_i)$, for every node $i$ of the PPI (normalization).
\end{enumerate}

In the experiments, we tested different choices for the Node Scoring Function $f$. The first set consists of functions that directly depend on $NR(i)$:
\begin{itemize}
    \item The \textbf{power scoring function}: $f(NR(i),\alpha) = NR(i)^\alpha$, with $\alpha > 0$\footnote{We considered $\alpha\in\{0.5, 1, 1.5, 2\}$}. When $\alpha=1$ we are directly using $NR(i)$ (default scoring function),
    \item The \textbf{sigmoid scoring function} outputs a value that is smooth and bounded based on two parameters: the steepness and translation parameters:
            $f(NR(i),s,\theta) =\frac{1}{ 1 + e^{-((NR(i) - \theta)\cdot s)}}$
\end{itemize}

We further considered node scoring functions that depend on the rank of PPI nodes in descending order of their values of $NR(i)$ (i.e., higher $NR(i)$, the lower the corresponding rank). In more detail, let $r_i$ denote the rank of node $i$. We considered the following, rank-dependent definitions for $f$:

\begin{itemize}
    \item \textbf{Linear}: Given the rank $r_i$ of node $i$ and the total number $N$ of nodes/proteins in the PPI, the linear ranking function is defined as $f(r_i) = \frac{N -r_i +1}{N}$.
    \item \textbf{Inverse Sigmoid:} In this case, for protein $i$ we have: $f(r_i) = 1 - \frac{1}{1 + e^{(-s \cdot (r_i - t))}}$.
\end{itemize}

\paragraph{Biological Transition Matrix (BTM)}
Though other choices are possible, for breast cancer, entry $W_{ij}$ of the random walk's transition matrix depends on the extent to which nodes $i$ and $j$ of the PPI share common annotations (i.e., they are involved in common biological processes) that are also significant for the disease. For breast cancer, we considered the following \textbf{Disease Specific Interaction} function:
\[
    DSI(i, j) = \frac{|enriched \_ set \cap label(i)\cap label(j)|}{|enriched \_ set|}.
\]
Intuitively, $DSI(i, j)$ will be higher, the more $i$ and $j$ share annotations that are also statistically significant for the disease under consideration.

$W_{ij}$ then depends on $DSI(i, j)$ according to a scoring function as follows:
\[
    W_{ij} = \begin{cases}
    f(DSI(i,j)) & \text{ if edge (i,j) belongs to PPI } \\ 
    0 & \text{ otherwise} 
\end{cases}
\]

In the experiments tested different choices for the scoring function $f$:
\begin{itemize}
    \item \textbf{Power scoring function}: for each edge $(i,j)$ we consider
        $f(DSI(i,j),\alpha) = DSI(i,j)^\alpha$, with $\alpha > 0$.
    \item \textbf{Summation scoring function}: for each edge $(i,j)$, we let
        $f(DSI(i,j),c) = DSI(i,j) + c$.
    \item \textbf{Sigmoid scoring function}: for each edge $(i, j)$, we have
        $f(DSI(i,j),s,t) =\frac{1}{ 1 + e^{-((DSI(i,j) - \theta)\cdot s)}}$, with $\theta$ and $s$ respectively the translation and steepness parameters.
\end{itemize}
\subsection{Internal validation}
\paragraph{Experimental setup}
For each algorithm and for each set of parameter values we considered, we considered the average value of $Recall@k$ (defined below) over $100$ independent runs. In each run, the seed set of known disease genes was randomly split into a \textit{training set}, accounting for $70\%$ of the original seed set, and a \textit{test set}, including the remaining $30\%$ of the genes.\footnote{For breast cancer, this amounts to $28$ and $12$ genes respectively.}
\paragraph{Performance measure}
Intuitively, we are interested in algorithms that identify new candidate genes that are more likely to be of interest for further biological scrutiny. Consistently, we measured performance using \textit{Recall}. This is the fraction of relevant items (in our case, known disease genes in the test set) that are successfully retrieved by the algorithm. Formally, in our scenario recall is defined as:
\[
    Recall = \frac{|disease \_ genes \cap retrieved \_ genes| }{|disease \_ genes|},
\]   
where $disease\_genes$ are known genes involved in the phenotype and $retrieved\_genes$ are genes prioritized by the algorithm under consideration.
Moreover, in order to compare our approach with other baselines, we considered  $Recall@k$. In our framework, this is the value of recall when $retrieved \_ genes$ the set of \textit{Top-K} genes in the algorithm's ranking. We considered several values for $k$, namely, $k\in \{10, 20, 40, 50, 60, 70, 80, 90, 100, 150, 200\}$.
\paragraph{Parameter setting}
As for DIAMOnD, this is a parameter-free algorithm. For RWR, we adopted the parameter setting suggested in  \cite{kohler2008walking}.

For BRW, as the previous paragraphs highlight, we used a grid search to select the parameter configuration used in the final round of experiments. In more detail, for each considered parameter configuration, we took the average value of $Recall@k$ over 1000 independent runs of BRW. The final configuration was the one achieving the best (average) $Recall@200$ score, \textit{e.g.} the ranking-inverse Sigmoid with steepness 0.01 and translation 250 for the BTP construction and the summation function for the DSI, with $c=1$.